\pdfoutput=1
\documentclass[12pt]{iopart}

\usepackage[colorlinks=true, citecolor=blue, urlcolor=blue ]{hyperref}
\usepackage{iopams}
\usepackage{slashed}
\begin{document}
\title{Lorentz Invariant CPT Violation for a Class of Nonlocal Thirring Model}

%&&&&&&&&&&&&&&&&&&&&&&&&&&&&&&&&&&&&&&&&&&&&&&&&
\author{Pinaki Patra}

\address{Department of Physics, University of Kalyani, India-741235}
\ead{monk.ju@gmail.com}
\author{Jyoti Prasad Saha}

\address{Department of Physics, University of Kalyani, West Bengal, India-741235}
\ead{jyotiprasadsaha@gmail.com}

%&&&&&&&&&&&&&&&&&&&&&&&&&&&&&&&&&&&&&&&&&&&&&&&&&
\begin{abstract}
It is possible to construct Lorentz invariant CPT violating models for Nonlocal Quantum Field Theory. In this article, we  present a class of  Nonlocal Thirring Models, in which the CPT invariance is violated while the Lorentz invariance is present. As a result, in certain cases the mass-splitting between particle and antiparticle are identified.
\end{abstract}


%&&&&&&&&&&&&&&&&&&&&&&&&&&&&&&&&&&&&&&&&&&&&&&&&&
\maketitle
\section{Introduction}
During mid period of 20'th century the nonlocal field theories \cite{Marnelius,Yukawa} were of great interest among the theoretical physicist and mathematicians (in this article by means of the nonlocal field theory we have considered the Lagrangian density where nonlocal interaction terms are present). The concrete mathematical formulation and as well as physical implications were studied in great detail in the literature of theoretical physics by several authors \cite{Yukawa,HYukawa,Qui,Soloview}. It was well known that within the Lorentz invariant framework every local field theory will obey the well established CPT theorem. For that reason, people often refer that the symmetry behind the CPT invariance is the Lorenz invariance. Also, it is an experimentally well established fact that the masses of particles and antiparticles are equal and it is a long time believe that this equality of masses for the particles and corresponding antiparticles is due to the CPT invariance of the theory. \\
For nonlocal theory the notion of particle and corresponding anti-particle is not in the position of clear understanding. Even if we adopt the similar meaning of particle and anti-particle  for the nonlocal field theory the situation will be very much different than that of local field theories. Recently, Chaichian et. all \cite{Chaichian} has constructed a class of nonlocal Lorentz invariant field theories which do not respect the CPT invariance principle. In \cite{Masud,Kazuo,Tureanu} they had also proved the possibility of mass-splitting for the non-local interaction term and they had argued that the equality of masses of particles and antiparticles is due to Lorentz invariance rather than to CPT. Recently, the recent data analysis \cite{Anirban,Aaltonen,Louvot,Aaij} speculates the neutrino antineutrino mass splitting which had drawn attention of Physicist to understand the main reason behind the mass splitting. It may be because of the effect of non-local interaction present in the process. Therefore, it is always interesting to study some non-local field theories; especially in the framework of CPT violation. One debatable issue is that whether the non-local theory may have a sensible S-matrix. Probably it was a letter of Yukawa \cite{Hideki} in which he first mentioned the S-matrix structure for the non-local theory ; but, that was critisized by Yennie \cite{Yennie} by showing that for general case of non-local field theory the convergence of S-matrix was ambiguous. Later, several works by Ruijsenaars\cite{Ruijsenaars} , Kuryshkin and Zorin \cite{Zorin} made the concrete S-matrix structure for non-local field theory.
Today probably it will not be a over ambitious claim if one say that " local and non-local fields may share the same S-matrix"\cite{Michael}. In this article we shall not distract ourselves by those controversy issues which are still on a shaky ground. Rather, we shall try to observe the possible particle anti-particle mass-splitting in Lorentz invariant CPT violating framework.\\
The main idea behind the process is to introduce an infrared divergent form factor \cite{Chaichian,Masud} in the interaction term. One can use the form factor $F((x-y)^2)\theta(x^0-y^0)\delta((x-y)^2)$; where $\theta(x)$ is the step function ($\theta(x)$  is equal to 1 for $x>0$ and $\frac{1}{2}$ for $x=0$ and vanishes otherwise). The weight function $F((x-y)^2)$ may be taken as for example Gaussian type $F((x-y)^2)= \exp\left(-\frac{(x-y)^2}{l^2}\right)$; $l$ being the non-locality length in the considered theory \cite{Chaichian}. In the similar manner we have inserted the infrared diverging form factor in the nonlocal interaction term and have studied the Thirring model in the aid of mass splitting of particle and corresponding antiparticle. \\
One may think that if one inserts the nonlocal interaction term, for every cases, the mass-splitting between particle and antiparticle will be observed; but, this is not the case. in this article by specific example we have shown that only specific choice of nonlocal interaction the mass-splitting of particle and antiparticle may be formulated.\\ The Thirring model is a completely soluble, covariant $1+1$- dimensional quantum field theory of a two-component Dirac spinor
\cite{Coleman,Fernandez,Alhaidari,Kaul,Acosta}. One important observation for the Thirring model is though the $Sine$-Gordon equation is the theory of massless scalar field, the $Sine$- Gordon soliton \cite{Coleman} can be identified with the fundamental fermion of the Thirring model.\\ 
In the next section we have studied three possible cases of non-local interaction term which follows by the conclusions.
\section{Non-local Models under Consideration}
The lagrangian for the well known local Thirring model is
\begin{equation}\mathcal{L}=i\bar{\psi}(x)\slashed{\partial}\psi(x) - m\bar{\psi}\psi + \lambda j^\mu(x)j_\mu(x)
\end{equation}
$j^\mu(x)=\bar{\psi}(x)\gamma^\mu\psi(x)$. \\
The notation $\slashed{A}$ stands for $\gamma^\mu A_\mu$.  m and $\lambda$ are mass and coupling constant respectively and $\gamma^\mu$ are the usual Dirac matrices. We shall be using the metric $g^{00}=-g^{11}=-1$. The $2\times 2$ Dirac matrices $\gamma^\mu$ obey the usual Clifford algebra,
\begin{equation}
\{\gamma^\mu,\gamma^\nu\}=2g^{\mu\nu} \nonumber
\end{equation}
We choose the representation via Pauli matrices,
$\gamma^0=\sigma_2$, $\gamma^1=i\sigma_1$, $\gamma_5=i\sigma_3$. \\
To study the non-local theory, we have chosen following three nonlocal models for our discussion  -\\
{\bf Case-1:}
\begin{eqnarray}
\mathcal{L}_{nonlocal}=-i\mu\int d^2xd^2y \bar{\psi}(x)\gamma^\mu\psi(y)\theta(x^0-y^0) \nonumber \\ \delta((x-y)^2-l^2)j_\mu(y) + C.C
\end{eqnarray}
The Lorentz invariance is manifested with the help of the $\theta$ functions introduced in the form factor. \\
The combination of charge conjugation, parity and time reversal, called CPT effectively flips the sign of all coordinates and performs a complex conjugation.   Non-conservation of CPT can be identified by straightforward manner i.e, by directly applying the CPT operation.  \\ 
The nonlocal interaction term transform under CPT as $-i\mu\int d^2xd^2y \bar{\psi}(x)\gamma^\mu\psi(y)\theta(y^0-x^0)  \delta((x-y)^2-l^2)j_\mu(y) + C.C $  which is not identical with the concerned nonlocal interaction term. \\
One should note that because of the scalar nature of the Lagrangian, the purely imaginary coupling ($i\mu$;with $\mu$ real) never appears for the local Lagrangian. But, for the nonlocal case we can introduce this one by the above-mentioned way. \\ Now, besides the case-1, we can consider the other two options to introduce the nonlocal terms as follows.\\

{\bf Case-2:}
\begin{eqnarray}
\mathcal{L}_{nonlocal}=-i\mu  \int d^2x d^2 y j^\mu(x) (\theta(x^0-y^0)-\theta(y^0-x^0))  \nonumber \\ \delta((x-y)^2-l^2)j_\mu(y) 
\end{eqnarray}

{\bf Case-3:}
\begin{eqnarray}
\mathcal{L}_{nonlocal}=-i\mu  \int d^2x d^2y \bar{\psi}(x)\gamma^\mu\psi(y)(\theta(x^0-y^0)- \theta(y^0-x^0)) \nonumber \\ \delta((x-y)^2-l^2) \bar{\psi}(x)\gamma_\mu\psi(y)
\end{eqnarray}
Direct applying CPT operation one can get that $\mbox{CPT}=-1$ for the last two types (case-2, case-3) of  interaction.
We will consider the total Lagrangian as the sum of the old local lagrangian and the $\mathcal{L}_{nonlocal}$ term. We will study each three cases separately. 
\section{Particle anti-particle mass-splitting}
\subsection{case-1}

The equation of motion for the nonlocal interaction term of case-1 reads
\begin{eqnarray}
i\slashed{\partial}\psi(x) - m\psi(x)+\lambda(\gamma^\mu\psi(x)j_\mu(x) + j^\mu(x)\gamma_\mu\psi(x))\nonumber \\ = i\mu  \int d^2y [\gamma^\mu\psi(y)\theta(x^0-y^0) j_\mu(y)- \gamma^\mu\psi(x)\theta(y^0-x^0) \nonumber \\ \bar{\psi}(x)\gamma_\mu \psi(y) - j^\mu (x)\theta(y^0-x^0)  \gamma_\mu \psi(y)]\delta((x-y)^2-l^2)
\end{eqnarray}
Now, if we put the ansatz $\psi(x)=U(p)e^{-ipx}$,  the above equation gives the dispersion relation
\begin{eqnarray}
\slashed{p}-m + \lambda(\gamma^\mu U(p)\bar{U}(p)\gamma_\mu + \bar{U}(p)\gamma^\mu U(p) \gamma_\mu ) \nonumber \\ 
= i\mu [\gamma^\mu U(p)\bar{U}(p)\gamma_\mu \{f_+(p)-f_-(p)\}-\bar{U}(p)\gamma^\mu U(p)\gamma_\mu f_-(p)]
\end{eqnarray}
Now, if we use the so called "Gordon identity" ($\bar{U}(p)\gamma^\mu U(p)\gamma_\mu=\frac{p^\mu}{m}\gamma_\mu$) and the completeness relation ($U(p)\bar{U}(p)=\slashed{p}+m$; spin sum is assumed), we can get the simplified forms 
\begin{eqnarray}
\gamma^\mu U(p)\bar{U}(p)\gamma_\mu= 4\gamma^\mu p_\mu -2m \\
\bar{U}(p)\gamma^\mu U(p) \gamma_\mu =\frac{1}{m}p^\mu \gamma_\mu
\end{eqnarray}
The choice of a light-like frame ($\vec{p}=0$) will enable us to write the following dispersive equation to determine the possible mass eigenvalues
\begin{eqnarray}
\gamma^0p_0 - m = -\lambda (4\gamma^0 p_0 -2m + \frac{1}{m} p^0 \gamma_0) + i\mu [(4\gamma^0p_0  \nonumber \\ -2m)(f_+(p^0)-f_-(p^0))-\frac{1}{m}p^0\gamma_0 f_-(p^0)]
\end{eqnarray}
In the above eigenvalue equation if we transform  $p_0\rightarrow -p_0$ and  sandwich the equation with the Chirality operator  $\gamma^5$  we get,
\begin{eqnarray}
\gamma^0p_0 - m = -\lambda (4\gamma^0 p_0 -2m + \frac{1}{m} p^0 \gamma_0) - i\mu [(4\gamma^0p_0  \nonumber \\ - 2m )(f_+(p^0)- f_-(p^0)) + \frac{1}{m} p^0 \gamma_0 f_+(p^0)]
\end{eqnarray}
which is different from the equation for $p_0$ (except for the case $\mu=0$). That means, if $p$ is the solution of equation (10), then $-p$ will not be the solution. Therefore, particle and antiparticle satisfy different mass eigenvalue equations. So, the nonlocal term in the Lagrangian splits the masses of particle and corresponding antiparticle.

\subsection{case-2}
 From the discussion of the previous case it is not difficult to understand that the appearance of $f_\pm (p)$ breaks the equality of particle anti-particle mass. Therefore, if we consider the non-local theory in which the effects of $f_\pm (p)$ goes away, no mass-splitting will be observed. This is exactly what one can see for the  Case-2. The straightforward calculation as mentioned in the previous subsection, gives the dispersion relation
\begin{equation}
\gamma^0p_0 - m = -\lambda (4\gamma^0 p_0 -2m + \frac{1}{m} p^0 \gamma_0)
\end{equation}  
which remains unchanged if we change $p\rightarrow -p$ and sandwich the equation with $\gamma^5$ i.e, if $p$ is the solution of (16), then $-p$ is also the solution of that equation. That means, this type of Lorentz invariant CPT violating non-local interaction also respect the equality of mass of particle and antiparticles which is not the situation for the previous type of non-local interaction where mass-splitting of particle and antiparticle was evident.
\subsection{case-3}
 With the similar prescription as described above one will get the particle mass eigenvalue equation as follows:
\begin{eqnarray}
\slashed{p}-m + \lambda(\gamma^\mu U(p)\bar{U}(p)\gamma_\mu + \bar{U}(p)\gamma^\mu U(p) \gamma_\mu ) \nonumber \\ 
= i\mu [\gamma^\mu U(p)\bar{U}(p)\gamma_\mu +\bar{U}(p)\gamma^\mu U(p)\gamma_\mu ]\{f_+(2p)-f_-(2p)\}
\end{eqnarray}
which  leads the remarks for mass splitting between particle and antiparticle.
\section{Discussion}
The above analysis shows that not all Lorentz invariant CPT violating non-local interaction gives the mass-splitting of particle and corresponding antiparticle. From the mathematical point of view the appearance of the modified form factor is responsible for the present scenario (also, that was the reason for the previous claim of Chaichian et. all.). As it is also possible to construct the CPT invariant Lotentz invariance violating models for the Noncommutative geometry, it is not quite clear what is the main reason for the equality of masses of particle and anti-particles. After all, besides some signature of neutrino-antineutrino mass-splitting (which is also rather small if one take the global average), there is no evidence of particle antiparticle mass-splitting. At, the same time one cannot ruled out the possibility of nonlocal interactions in a process. So, we can argue that the main reason behind the equality of particle antiparticle mass (for almost all the experimentally verified cases) is not quite well understood.\\
It may be noted that the cases for which the mass-splitting were observed were not local gauge-invariant. Besides, mass-splitting was not observed in the case-3 which was U(1) local gauge-invariant. \\
Furthermore, it is well known that the inclusion of non-locality may breaks the unitarity of the theory in general. In our case we take our non-local coupling ($\mu$) small such that the violation of unitarity is minimal. \\
Therefore, one can conclude that within the minimal unitarity violating Lorentz-invariant CPT violating non-local theory the inclusion of infrared divergent form factor may breaks the equality of masses of particles and corresponding antiparticles.\\
 If one intends to handle the situation in the scenario of gauge-invariance, one can insert a swinger non-integrable phase factor in the nonlocal term and one possible way to handle the non-integrable phase factor is to replace this by a propagation of very massive (indefinite) particle which makes the fermion line discontinuous \cite{ATureanu}. The possible scenario for the non-local Thirring model coupled with electromagnetic field in gauge invariant manner will be discussed elsewhere by the present authors.

\section*{Acknowledgments :}We like to thank Prof. Masud Chaichain for discussion. Pinaki Patra is grateful to CSIR, Govt. of India for fellowship support which made this work possible. Jyoti Prasad Saha like to acknowledge the financial support under PURSE from DST, Govt. of India.

\end{document}